\documentclass[
 amsmath,amssymb,
 aps,
twocolumn,
letter
]{revtex4-2}
\usepackage[utf8]{inputenc}
\usepackage{amsmath}
\usepackage{graphicx}
\usepackage{appendix}
\usepackage{color}
\usepackage{hyperref}

\begin{document}
\title{Yang-Lee zeros for real-space condensation}
\author{Zdzislaw Burda}\email{zdzislaw.burda@agh.edu.pl} 
\affiliation{AGH University of Krakow, Faculty of Physics and Applied Computer Science, 
al. Mickiewicza 30, 30-059 Krak\'ow, Poland}
\author{Desmond A. Johnston}\email{D.A.Johnston@hw.ac.uk}
\affiliation{School of Mathematical and Computer Sciences, 
Heriot-Watt University, Riccarton, Edinburgh EH14 4AS, UK }
\affiliation{Maxwell Institute of Mathematical Sciences,
Edinburgh, UK}
\author{Mario Kieburg}\email{m.kieburg@unimelb.edu.au}
\affiliation{University of Melbourne, 
School of Mathematics and Statistics, 
Parkville, VIC, 3010,
Australia} 
\affiliation{AGH University of Krakow, Faculty of Physics and Applied Computer Science, 
al. Mickiewicza 30, 30-059 Krak\'ow, Poland}
\affiliation{Faculty of Physics, Bielefeld University, PO-Box 100131, D-33501 Bielefeld, Germany}

\begin{abstract}
Using the electrostatic analogy, we derive an {\it exact formula } for the limiting Yang-Lee zero distribution in the random allocation model of general weights. This exhibits a real-space condensation phase transition, which is induced by a pressure change. The exact solution allows one to read off the scaling of the density of zeros at the critical point and the angle at which locus of zeros hits the critical point. Since the order of the phase transition and critical exponents can be tuned with a single parameter for several families of weights, the model provides a useful testing ground for verifying various relations between the distribution of zeros and the critical behavior, as well as for exploring the behavior of physical quantities in the mesoscopic regime, i.e.,  systems of large but finite size. 
The main result is that asymptotically 
the Yang-Lee zeros are images of a conformal mapping, given by the generating function for the weights, of uniformly distributed complex phases.

\end{abstract}

\maketitle

\paragraph*{Introduction.}

Lee and Yang~\cite{LY} were the first to observe that the non-analyticity in the thermodynamic potential signaling a phase transition could be understood from the behaviour of the corresponding partition function zeros for finite-sized systems by using a suitably complexified fugacity. 
The appearance of a critical point 
was due to the accumulation of the zeros on the positive real axis of the complex fugacity plane, which approached this point from the upper and lower complex half-plane for increasing system size, ``pinching'' the real axis. This picture is quite general and applies to a whole range of statistical 
models in which a phase transition occurs in the thermodynamic limit~\cite{LY, Fi64, Abe, Suzuki, wjk}. 

Typically, the partition function zeros are extracted from series expansions and transfer matrix calculations or from numerical simulations of finite-sized systems~\cite{lottsashrock, lottsashrock2}. Only in a very limited number of cases it is possible to analytically determine the thermodynamic limit of the locus of zeros.  These include some one-dimensional~\cite{1d1, 1d2, 1d3} and mean-field type models~\cite{MF1}, hierarchical models~\cite{IPZ}, recursive models~\cite{BGP,prz}, the random energy model~\cite{REM}, a random matrix model in QCD~\cite{HJV1997}, the Ising model on planar random graphs~\cite{stau, fatfish}, and the classical two-dimensional Ising model~\cite{LY, Fi64}. Beyond those areas the literature becomes very scarce, which is understandable, as the mathematical problem about the mother body (as the support of the zeros is known in mathematics, also known as potential theoretic skeleton or S-curve) is highly nontrivial~\cite{bj,mfr}. Therefore, it is always of interest to find another family of physical and/or statistical models where such an explicit analysis is possible, as they can serve as blueprints for more involved systems.

In the present work, we have identified and investigated another analytically accessible example which is rather versatile and  simple. It is the random allocation model~\cite{Ritort,bbj,balls2}, whose origin lies in Ehrenfest's urn model~\cite{Ehrenfest} and which exhibits a condensation transition. This model has been applied in the past (under the moniker of the ``balls in boxes model''  or ``backgammon model'') to such  diverse phenomena as collapse transitions in branched polymers (random trees)~\cite{,bb}, studies of glassy and non-equilibrium systems of condensate formation~\cite{Ritort,fr,gl,rs03,gpr09}, diversity in  
population models~\cite{binns,div}, mass transport~\cite{mez,EW14},
phases in discrete models of quantum gravity~\cite{bbj2,rf15,br17} and even wealth condensation in econophysics~\cite{bjjknpz}
as well as network models~\cite{dm,mb08,janson12}, biology~\cite{ew07}, combinatorial problems in representation theory~\cite{b21} and number theory~\cite{h11,bevan}. We concentrate on the equilibrium systems but many of the results can be applied to non-equilibrium context.

In principle, the phase transition that occurs in this model can be used to describe the process of nucleation of fireball formation \cite{vH} or more generally the appearance of the giant component which is observed in many phenomena ranging from percolation \cite{as}, networks \cite{babi} to renewal processes \cite{g2}. 
The partition function of the model can also be thought of as arising from the non-equilibrium steady state 
of a zero range process 
with suitably chosen jump rates \cite{eh}. Although the model is well known, the corresponding Yang-Lee zeros have not been analytically computed previously. We show that this is indeed possible in a very general setting.

The random allocation model has several attractive features for studying the properties of partition function zeros. Once suitable weight functions are chosen, the phase transition can be tuned to first or any higher order by varying a single parameter. Usually logarithmic corrections to scaling appear for integer values of such parameters. Thus, one can employ those results as benchmarks. The zeros may also be calculated efficiently for finite-sized systems~\cite{bbjz} which facilitates comparison with the thermodynamic limit in investigations of the scaling properties. Finally, as we derive here, there is a particularly simple formula for the locus of zeros in the thermodynamic limit.

The method we use has deep physical roots, as it is based on two-dimensional electrostatic equations that relate the charge distribution (in this case the Yang–Lee zeros) and the electric field (in this case the complex derivative of the thermodynamic potential). This method has been proven 
to work in many different contexts, for example in random matrix theory where it is
used to derive eigenvalue statistics of various ensembles, e.g., see~\cite{HJV1997}. As we shall see, this method also turns out to be indispensable in finding the distribution of the zeros of the
random allocation model partition functions and in explaining the mechanism of the phase transition that takes place in the model.
 
\paragraph*{The Model.}
We consider a statistical system of $S$ particles distributed in $N$ boxes,
described by the partition function 
\cite{bbj, balls2, dgc, g}
\begin{equation}
    Z_{S,N} = {\sum}_{s_1,\ldots,s_N\geq1} 
    w(s_1) \ldots w(s_N) \delta_{S- (s_1+\ldots +s_N)} .
    \label{ZSN}
\end{equation}
The weights $w(s)$ are non-negative and defined on the set of positive integers, 
$s=1,2,\ldots$ If we view a box as an elementary volume element, then such a system 
can be understood as a gas of particles with zero-range interactions that occur only 
within the elementary volume elements. The zero-range interactions are entirely defined by the weight function $s \rightarrow w(s)$, corresponding to the statistical weight of packing $s$ particles into a single volume element. To simplify the notation, we assume that each volume element contains at least one particle, but the results can easily be translated via a shift to the case where there are also empty volume elements.

The weights can be encoded in the generating function
\begin{equation}
f(z) = {\sum}_{s=1}^\infty w(s) z^s=z {\sum}_{s'=0}^\infty w(s'+1) z^{s'}.
\label{f}
\end{equation}
Then, the partition function is given as the contour integral
\begin{equation}
Z_{S,N} = \oint_{|z|=\epsilon} \frac{dz}{2\pi i} \frac{f(z)^N}{z^{S+1}} ,
\label{oint}
\end{equation}
 with a suitably small radius $\epsilon>0$ such that the function $f(z)$ is holomorphic on $|z|<\epsilon$. 

The most interesting family of weights comprises those
for which the corresponding generating function series~\eqref{f} has a finite radius
of convergence $z_{\rm c}$, because in this case the system may undergo a real-space condensation phase transition \cite{bbj,s,e,jmp,kmh,emz2,gss,fls,cg2,jcg,mez1}. Due to a trivial rescaling $f(z)\to f(z_{\rm c}z)$, we can restrict 
ourselves to weights for which the radius of convergence of the series~\eqref{f})
is $z_{\rm c}=1$. 

An example is the family of power-law weights 
\begin{equation}
\label{plw}
w(q) = q^{-\beta},  \ {\rm for} \ q=1,2,\ldots
\end{equation}
for $\beta \in (1,\infty)$ for which the generating function (\ref{f}) is given by the polylogarithm $f(z) = {\rm Li}_\beta(z)$. In the thermodynamic limit $N \rightarrow \infty$ and $S/N \rightarrow \rho$ where $\rho$ is the limiting density averaged over all boxes, the system undergoes a phase transition at the critical value 
$\rho_{\rm c} = f'(1)/f(1)$ (if its limit $z\nearrow1$ is finite). For $\rho>\rho_{\rm c}$ a finite fraction of particles condenses in a single volume element (box). For the power-law weights the critical
density is $\rho_{\rm c} = \zeta(\beta-1)/\zeta(\beta)$ for $\beta>2$, where $\zeta$ is the Riemann zeta function.  This is the reason why this model with these specific weights is called the zeta-urn model~\cite{gl.zeta.urn}.
The order of the transition changes monotonously from infinite to third order when $\beta$
increases in the range  $(2,3)$.  
The transition is second order for $\beta \in [3,\infty)$.

\paragraph*{Yang-Lee Zeros.}
We focus on the partition function for the isobaric (elsewhere denoted as ``grand-canonical'')  ensemble~\cite{bbj2, balls2, g3, bbjr}
\begin{equation}
    Z_{S,\mu} = {\sum}_{N=1}^S  Z_{S,N}e^{-\mu N}={\sum}_{N=1}^S Z_{S,N} u^N=Z_{S}(u).
    \label{ZSmu}
\end{equation}
with pressure $\mu$ and fugacity $u=e^{-\mu}$. Our interest is the (Yang-Lee) zeros $\xi_S(j)$ of these polynomials~\cite{bbjz}, which also admit the form
\begin{equation}
Z_S(u) = w(1)^S {\prod}_{j=1}^S (u - \xi_S(j))
\label{ZS}
\end{equation}
with
$w(1)=Z_{S,N=1}$. The polynomial has no constant term, so it has one trivial zero at $u=0$. Since all coefficients of the polynomial $Z_S(u)$
\eqref{ZSmu} are non-negative (as is normally the case in statistical mechanical models), there are no zeros on the positive real axis $(0,\infty)$.
Zeros occur in complex conjugate pairs, or lie on $(-\infty,0]$. 
Hence, the function 
\begin{equation}
\psi_S(u) = \frac{\ln Z_S(u)}{S} =  \ln w(1) + \frac{1}{S} {\sum}_{j=1}^S \ln (u - \xi_S(j))
\label{psiSu}
\end{equation}
is analytic about any positive $u=e^{-\mu}>0$ for any finite $S$.
Following the Lee-Yang methodology, the mechanism of creating a critical point $u_{\rm c} = e^{-\mu_{\rm c}}$ on the positive real axis 
where the thermodynamic potential
\begin{equation}
\widehat{\psi}(\mu) = \lim_{S\rightarrow \infty} \psi_S(e^{-\mu}) 
\end{equation}
has a singularity in the thermodynamic limit, is related to the accumulation of zeros at 
$u_{\rm c}$ which approach this point from the upper and lower complex half-plane as $S$ increases.

Defining the density of zeros on the complex plane
\begin{equation}
    \varrho_S(\xi) = \frac{1}{S} {\sum}_{j=1}^S \delta(\xi - \xi_S(j))
\end{equation}
we can write~\eqref{psiSu} as 
\begin{equation}
    \psi_S(u) = \psi_0  + \int d^2 \xi \varrho_S(\xi) \ln (u - \xi) 
    \label{psiS}
\end{equation}
where $\psi_0 = \ln w(1)$. Using an electrostatic analogy~\cite{bbjz, BDL, Brazil}, this equation can be interpreted as a relationship between the charge density $\varrho_S(z)$ and the electrostatic potential ${\rm Re}(\psi_S(u))$ in two dimensions.
Taking the derivative with respect to $u$ gives us the equation for the electric field produced by the electric charges (located at zeros of the partition function)
\begin{equation}
    \psi'_S(u) = \int d^2 \xi \frac{\varrho_S(\xi)}{u - \xi} .
    \label{fpS}
\end{equation}
In the thermodynamic limit, $S\rightarrow \infty$, the finite-$S$ functions are replaced with the limiting functions $\psi_S(u) \rightarrow \psi(u)$ and $\varrho_S(u)\rightarrow \varrho(u)$. In other words,
from the locus of zeros we can predict the critical behavior of
the statistical system in the thermodynamic limit, and the way it
is generated from mesoscopic systems of large but finite size $S$.

\paragraph*{The Thermodynamic Limit.}
Substituting~\eqref{oint} into~\eqref{ZSmu}, the sum over $N$ can be carried out to yield
\begin{equation}
    Z_S(u)=\oint_{|z|=\epsilon} \frac{dz}{2\pi iz}\frac{f(z)u-[f(z)u]^{S+1}}{z^{S}(1-f(z)u)}.
    \label{ZS_geom}
\end{equation}  
The series $f(z)$ does not have a constant term, so $f(z)/z$ is also holomorphic inside the disk and the integral over the second term with $[f(z)u]^{S+1}$ evaluates to zero if $\epsilon>0$ is small enough for $f(z)u=1$ to have no solution inside the contour. Expressing the remaining numerator as $f(z)u=f(z)u-1+1$, we arrive at the simplification
\begin{equation}
    Z_S(u)=\oint_{|z|=\epsilon}\frac{dz}{2\pi iz^{S+1}}\frac{1}{1-u f(z)}.
    \label{uzeros}
\end{equation}
This expression opens the path to the limiting density $\varrho(\zeta)$ via $\psi_S(u)$, see~\eqref{psiSu}.

The standard approach is to calculate $\psi'_S(u)$ for a given density $\varrho_S(\xi)$ via~\eqref{fpS}. Here, we are considering the inverse problem of calculating $\varrho_S(\xi)$ (and thereafter the critical behavior) for a given $\psi'_S(u)$, which in general is  more difficult  than the standard approach. This inverse problem simplifies significantly, however,  when the locus of zeros forms a one-dimensional curve (the {\it accumulation curve}) $\gamma$
in the limit $S\rightarrow \infty$. In this case,~\eqref{fpS} takes the form
\begin{equation}
    \psi'(u) = \lim_{S \rightarrow \infty} \bigl( \psi'_S(u) - \frac{1}{S u} \bigl) = 
   \int_\gamma d\xi \frac{\varrho(\xi)}{u-\xi},
   \label{psi}
\end{equation}
where
the pole $1/u$ coming from the trivial zero of $f(z)$ at $z=0$ is subtracted.
This simplification occurs in many statistical mechanical systems. 
The idea is to exploit the non-analyticity of $\psi'(u)$ along $\gamma$ in~\eqref{psi} and employ the Sokhotskii-Plemelj formula
\begin{equation}
  \label{rho}
 \varrho(\gamma(t))= \frac{1}{2\pi i}\lim_{\delta\to 0^+}\big[\psi'(\gamma(t)+n(t)\delta)- 
 \psi'(\gamma(t)-n(t)\delta)\big],
\end{equation}
where $\gamma$ is parameterized by $t\rightarrow \gamma(t)$ and the unit normal vector to the curve at the point $\gamma(t)$ is $n(t)$. 

To derive the accumulation curve $\gamma$ from~\eqref{uzeros}, we need two ingredients. One is the radius of convergence $z_{\rm c}=1$, another  is Alexander's condition~\cite{Alexander} 
\begin{equation}\label{alex}
{\sum}_{q=1}^\infty |qw(q)-(q+1)w(q+1)|\leq 1  .
\end{equation}
for the weights, which we assume here.
It is easy to check that power-law weights fulfil this. Equation~\eqref{alex} implies that $f(z)$ is injective within the unit disc $D=\{z:|z|< 1\}$, meaning it is a conformal map, so that $1-u f(z) =0$ has a unique solution
\begin{equation}
z=z_0(u)=f^{-1}(1/u)
\label{pole}
\end{equation} 
when $u^{-1}\in f(D)$. The accumulation curve is then simply $\gamma=1/f(\partial D)$ as we will show.

For large $S$ the high order pole at the origin pushes all saddle points of the integrand away from $z=0$. When $u^{-1}\in f(D)$ is in the exterior of $\gamma$,  the only saddle point $z_{\rm SP}$ inside the disc of convergence $D$ is very close to the pole~\eqref{pole} which is derived from the saddle point equation
\begin{equation}\label{sadd.eq}
    \frac{S+1}{z_{\rm SP}}=\frac{u f'(z_{\rm SP})}{1-uf(z_{\rm SP})}
\end{equation}
Indeed, $z_{SP}$  expanded in $1/S$ is
\begin{equation}
   \label{deviation}
    z_{\rm SP}=z_0(u)\left[1-S^{-1}+\mathcal{O}\left(S^{-2}\right)\right],
\end{equation}
where $f(z_{\rm SP})=f(z_0(u))-f'(z_0(u))z_0(u)/S+\mathcal{O}(S^{-2})$.
Since $z_{\rm SP}\in D$, too, we can choose the radius $\epsilon=|z_{\rm SP}|$ by Cauchy's theorem. 
For large $S$ the integral, then, 
acquires its main contribution from the neigbourhood of $z_{\rm SP}$ so that we expand $z=z_{0}(u)(1- [1-i\delta z]/S)$ yielding an integral over $\exp[1-i\delta z]/(1-i\delta z)$ along the real line for $S\to\infty$. Integrating over $\delta z$ 
results in
\begin{equation}
    Z_S(u) \sim \frac{1}{uf'(z_0(u))[z_0(u)]^{S+1}},
\end{equation}  
so we eventually obtain
\begin{equation}
   \lim_{S\rightarrow \infty}\frac{1}{S}\ln Z_S(u) = \psi(u) = - \ln z_0(u). 
\end{equation}

The situation changes drastically for $u\notin 1/f(D)$ in  the interior of $\gamma$. Then,
the term $1/(1-u f(z))$ has no pole inside the unit disc and the maximal possible
deformation of the integration contour which does not change the integral 
is its enlargement to the circle with the radius of convergence $z_{\rm c}=1$. We assume a finite number of branch points of $f(z)$ at $z_1,\ldots,z_L\in\partial D$, where $z_1=1$ is always one of them due $|f(z)|\leq f(|z|)$ and the radius of convergence $z_{\rm c}=1$. The reality of the coefficients of $f(z)$ also implies that branch points come in complex conjugate pairs. As the number $L$ of branch points is finite  we can enlarge the circle even further beyond the radius $z_{\rm c}=1$ but need to walk around the branch cuts via the paths $z=z_j(1+[x\pm i\delta]/S)$ with $\delta\to0$. The integral over the circle is exponentially suppressed due to the term $1/z^{S}$ and a radius $r>1$ that we choose $r-1\ll 1/\sqrt{S}$ while the branch cuts give the main contribution,
\begin{equation}
    \begin{split}
        &Z_S(u)\sim{\sum}_{j=1}^L\frac{1}{\pi S\,z_j^{S}}\lim_{\delta\to0}\int_0^\infty dx e^{-x}\\
        \times&\left[\frac{e^{i\delta}}{1-uf(z_j(1+\frac{x+ i\delta}{S}))}-\frac{e^{-i\delta}}{1-uf(z_j(1+\frac{x- i\delta}{S}))}\right].
    \end{split}
\end{equation}
This expression shows that $Z_S(u)$ can vanish at most algebraically in $S\to\infty$ while it is bounded from above so that
$\ln Z_S(u)/S \rightarrow \psi(u) = 0$
for $S\rightarrow \infty$ and $u\notin 1/f(D)$. We believe that this result will be even true for cases where the number of branch points of $f$ on the unit circle is infinitely large though the argumentation above must be modified.

In summary, the solution consists of two parts 
\begin{equation}
   \label{psi_solution}
    \psi(u)=\left\{\begin{array}{cc} - {\rm ln}z_0(u), & u\ {\rm in\ the\ exterior\ of}\ \gamma,\\
    0, & u\ {\rm in\ the\ interior\ of}\ \gamma. \end{array}\right.
\end{equation}
In the thermodynamic limit, $\psi(u)$ thus
depends only on the generating function~\eqref{f} via $z_0(u) = f^{-1}(1/u)$, see~\eqref{pole}, so that $\gamma$ is the image $1/f(\partial D)$ of the unit circle.

The density of zeros along $\gamma$ can be found using~\eqref{rho},
\begin{equation}
\label{rho_u}
    \varrho(u)du=\frac{z'_0(u)}{2\pi i z_0(u)}du=\frac{1}{f^{-1}(1/u)f'\left(f^{-1}(1/u)\right)}\frac{d(1/u)}{2\pi i}.
\end{equation}
Choosing the parameterisation $u=1/f(e^{i\varphi})$, for which the Jacobian is $d(1/u)=ie^{i\varphi}f'(e^{i\varphi})d\varphi$, we obtain
\begin{equation}
\label{rho_phi}
    \varrho(u)du=d\varphi/2\pi. 
\end{equation}
Thence, $\varrho(u)$ is just the image of the {\it uniform} density on the unit circle 
under the conformal map $u=1/f(z)$. 

The critical value of the pressure $\mu=-\log(u)$ is given by $\mu_{\rm c}=\log f(1)$ which shows that for a finite critical value we need a finite $f(1)<\infty$. In comparison, the canonical ensemble with a finite critical density $S/N\to\rho_{\rm c}=f'(1)/f(1)$  requires a finite first derivative $f'(1)$ which also implies a finite $f(1)$. Thus, a phase transition in the canonical ensemble implies one in the isobaric, but the converse is not true. This was already pointed out in~\cite{balls2}.

\begin{figure}[t!]
    \centering
    \includegraphics[height=0.25\textwidth]{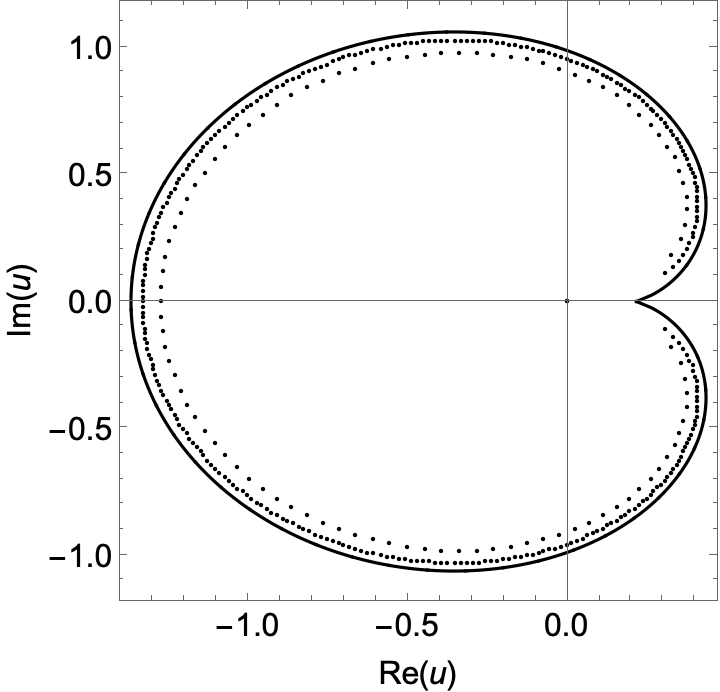} 
    \caption{Zeros $\xi_S(j)$ (\ref{ZS}) for $S=100$ and
    $S=300$ for the power law weights with $\beta=5/4$ compared to the theoretically calculated accumulation curve 
    $\gamma=1/ f( \partial D) = 1/{\rm Li}_\beta(e^{i\varphi})$.
    \label{fig:th_vs_num}}
\end{figure}

\paragraph*{Examples.}
For illustration, consider the power-law weights~\eqref{plw} where $f(z) = {\rm Li}_\beta(z)$. In the isobaric ensemble, the phase transition is of order $1 + \lfloor 1/(\beta-1) \rfloor$. It is therefore first order for $\beta \in (2,\infty)$ and changes from second to infinite order as $\beta$ is reduced from two to one in the range $\beta \in (1,2]$.  The critical pressure is $\mu_{\rm c}=\log \zeta(\beta)$. In Fig.~\ref{fig:th_vs_num} we show the locus of zeros for
finite $S$ that tend to $\gamma$ as $S$ increases.
The angle at which $\gamma$ crosses $u_{\rm c}=e^{-\mu_{\rm c}}$ depends on $\beta$ and can be determined
from the expansion~\cite[25.12.12]{NIST} (for non-integer $\beta$, for integer $\beta$ there are additional logarithmic singularities \cite{balls2}),
\begin{equation}
  \label{Li}
  {\rm Li}_\beta(e^{i\varphi}) = \Gamma(1-\beta) (-i\varphi)^{\beta-1} + {\sum}_{k=0}^\infty 
  \frac{\zeta(\beta-k)}{k!} (i\varphi)^k.
\end{equation}
For $\beta\in (1,2)$, it is
\begin{equation}
\Delta u = u - u_{\rm c} = c e^{- i (\beta-1) \pi\varphi/2|\varphi|} |\varphi|^{\beta-1} + o(\varphi^{\beta-1}),
\end{equation}
where $c= -\Gamma(1-\beta)/\zeta^2(\beta)>0$ so that $\gamma$
crosses $u_{\rm c}$ with the angle $(\beta-1) \pi/2$. 
The density of zeros near $u_{\rm c}$ as a function of
the distance $|\Delta u| \sim \phi^{\beta-1}$ behaves as $\varrho(|\Delta u|) \propto |\Delta u|^{1/(\beta-1) -1}$ when $|\Delta u| \rightarrow 0$ due to~\eqref{rho_u}. 

For $\beta>2$ (the first order phase transitions)
the leading behaviour for $\varphi \rightarrow 0$ 
originates from the term for $k=1$ in the sum (\ref{Li}) and not from the term $(-i \varphi)^{\beta-1}$,
which becomes sub-leading in this case. As a consequence, the accumulation curve $\gamma$ crosses $u_{\rm c}$ perpendicularly and the density approaches a positive constant for $|\Delta u|\rightarrow 0$. 

\begin{figure}[t!]
    \centering
    \includegraphics[height=0.25\textwidth]{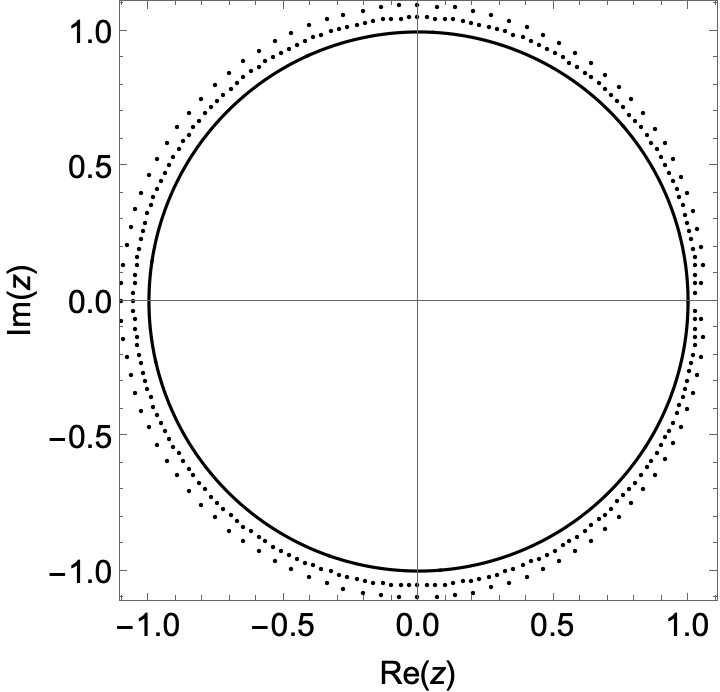} 
    \caption{Sets of preimages $z_j=f^{-1}(1/\xi_S(j))$ of zeros (\ref{ZS}) for binomial weights with $\theta=1/2$, for $S=100,200$ (dots). The points $z_j$ are approximately uniformly distributed on the circles which tend to the unit circle (solid line) for $S\rightarrow \infty$.
    \label{fig:preimage}}
\end{figure}

As a second example, we consider the family of binomial weights $w(q) = (-1)^{q+1} \binom{\theta}{q}$ for $\theta \in (0,1)$ and $q=1,2,\ldots$  
The generating function~\eqref{f} as well as its inverse take the explicit forms
\begin{equation}
    f(z) = 1 - (1-z)^\theta\quad {\rm and}\quad  f^{-1}(u) = 1 - (1-u)^{1/\theta}.
\end{equation}
The phase transition is at $e^{-\mu_{\rm c}}=u_{\rm c} = f(1)=1$ and is of 
order $1 + \lfloor 1/\theta \rfloor$. In the thermodynamic limit, $\psi(u)$ is given  by~\eqref{psi_solution} with $z_0(u)=f^{-1}(u)$. Since $f^{-1}(u)$ is explicitly known, we can  numerically compute the zeros $\xi_S(j)$ for large but finite $S$, and then use this inverse map to find their preimage 
$z_j(S) = f^{-1}(1/\xi_S(j)) = 1- (1-1/\xi_S(j))^{1/\theta}$. 
The results for $\theta=1/2$
are shown in Fig.~\ref{fig:preimage}.
Interestingly, the zeros still exhibit a very regular separation. This suggests that even at finite $S$ a conformal map exists so that the zeros become the roots of unity (times a possible shift in the complex phase).

\paragraph*{Conclusion.}
Our main conclusion is that for $S\to\infty$ the locus of zeros of the isobaric partition function $Z_S(u)$ in~\eqref{ZS} is simply the image $u=1/f(e^{i\varphi})$ with $\varphi\in[0,2\pi)$ uniformly distributed, where $f(z)$ is the generating function for the statistical weights~\eqref{f}. Critical properties of the system, such as the phase transition or the angle at which the accumulation curve hits the positive real axis, are straightforward consequences of this observation. 
Since both the thermodynamic limit and finite size results are accessible in the random allocation model, one obvious avenue for further exploration is to consider the integer powers which give rise to logarithmic scaling and compare in detail the scaling behavior with general expectations~\cite{log}.   
An open issue is how to modify the conclusion when the series~\eqref{f} is no longer injective in the disc of convergence, in particular when Alexander's condition~\eqref{alex} is not satisfied.  

Last but not least, the electrostatic analogy used in this article to find the distribution of zeros works so well because the zeros coalesce to form a one-dimensional curve in the limit. The question is if one can apply this method if the limiting distribution
is two-dimensional. The answer is affirmative: there is a variation of this method that works also for two-dimensional regions. The method was developed to determine the eigenvalue spectra of non-Hermitian random matrices and it is based on an extension of the Sokhotskii-Plemelj formula to quaternion space \cite{quat1,quat2}.

\begin{acknowledgments}
MK is funded by the Australian Research Council via the
Discovery Project grant DP210102887, by the  AGH University's visitor program and by the Alexander-von-Humboldt Foundation.
He also thanks the Faculty of Physics at the AGH University, the Faculty of Physics at Bielefeld University and of the Centre for interdisciplinary Research  ZiF for the hospitality during his sabbatical.
\end{acknowledgments}

\end{document}